# Deep Transfer Learning Based Intrusion Detection System for Electric Vehicular Networks

Sk. Tanzir Mehedi [1], Adnan Anwar [2,*], Ziaur Rahman [1] and Kawsar Ahmed [1]

[1] Department of Information and Communication Technology, Mawlana Bhashani Science and Technology University, Tangail 1902, Bangladesh; tanzirmehedi@ieee.org (S.T.M.); zia@iut-dhaka.edu (Z.R.); k.ahmed.bd@ieee.org (K.A.)
[2] Centre for Cyber Security Research and Innovation (CSRI), Deakin University, Geelong 3216, Australia,
\* Correspondence: adnan.anwar@deakin.edu.au; Tel.: +61-3-522-73679

**Abstract:** The Controller Area Network (CAN) bus works as an important protocol in the real-time In-Vehicle Network (IVN) systems for its simple, suitable, and robust architecture. The risk of IVN devices has still been insecure and vulnerable due to the complex data-intensive architectures which greatly increase the accessibility to unauthorized networks and the possibility of various types of cyberattacks. Therefore, the detection of cyberattacks in IVN devices has become a growing interest. With the rapid development of IVNs and evolving threat types, the traditional machine learning-based IDS has to update to cope with the security requirements of the current environment. Nowadays, the progression of deep learning, deep transfer learning, and its impactful outcome in several areas has guided as an effective solution for network intrusion detection. This manuscript proposes a deep transfer learning-based IDS model for IVN along with improved performance in comparison to several other existing models. The unique contributions include effective attribute selection which is best suited to identify malicious CAN messages and accurately detect the normal and abnormal activities, designing a deep transfer learning-based LeNet model, and evaluating considering real-world data. To this end, an extensive experimental performance evaluation has been conducted. The architecture along with empirical analyses shows that the proposed IDS greatly improves the detection accuracy over the mainstream machine learning, deep learning, and benchmark deep transfer learning models and has demonstrated better performance for real-time IVN security.

## 1. Introduction

In recent years, the automotive industry has been undergoing a radical transformation. With the ongoing development of network communication, modern vehicles are rapidly transitioning from fully mechanical to software-controlled technologies [1]. Modern In-vehicle Network (IVN) technologies and services are being integrated with intelligent information systems. As a result, the number of IVN devices is rapidly increasing and becoming more complex. The IVN devices must be seamlessly connected to an external network system in order to receive communication services efficiently. However, this increases the risk of the IVN to potential internal or external threats. The Electronic Control Units (ECUs) are software-controlled technologies that read various sensor data and perform relevant processing, including automatic brake control, pedestrian detection, auto-parking, path-planning, actuators control, and collision avoidance [2]. The sensor and actuator values are transmitted to other ECUs via the IVN protocol, resulting in the formation of a very complex network. There are several IVN protocols in the automotive industry, including Controller Area Network (CAN), Controller Area Network Flexible Data-Rate (CAN FD), Media Oriented Systems Transport (MOST), FlexRay, and Local Interconnect Network (LIN) [3]. Among all the data communication buses, CAN bus is the most well-known and extensively used protocol in the automotive vehicles industry [4]. Furthermore, the CAN buses are being applied also in other industries, including agriculture, aerospace, medical devices, and commercial machinery [1]. Several other protocols are also available with more security features (e.g., Ethernet). However, in the field of automotive IVN communication, these advanced protocols can not be completely replaced by the CAN bus protocol due to some reasons [4]. First of all, the CAN bus is more design flexible and perfectly appropriate for real-time environments, ensuring secure and fast communication between ECUs with minimal latency time. Secondly, there is a process of prioritization in the CAN bus protocol that prevents lower-priority messages from interfering with higher-priority messages. To cite an example, a message that transmits a more critical function such as an engine control message takes precedence over a door control message. Finally, the CAN bus protocol serves as the backbone of automotive IVN communication in all modern vehicles. To completely replace this protocol with another, the IVN architecture must be completely redesigned. As a result, other protocols will not completely replace the CAN bus's role and application.

However, in-vehicle intrusion detection has become a growing interest field that has been researched across a wide range of disciplines. In the CAN bus protocol, intrusion detection is the method of monitoring normal and abnormal traffic between different ECUs and identifying any abnormal traffic using Traditional Machine Learning (TML) algorithms [5]. With the rapid development of IVNs and evolving threat types, the TML-based IDS has to update to cope with the security requirements of the current environment. Nowadays, regarding the progression of Deep Learning (DL), Deep Transfer Learning (DTL), and its impactful outcome in several areas, these techniques have gained the attention of many researchers in the field of cybersecurity (e.g., IDS, antivirus or malware identification) [6]. In particular, in the field of the automotive industry, recent DL and DTL-based IDS have also gained the attention of many researchers, which is discussed further in the Related Work section.

Automobile manufacturers are working to develop fully autonomous vehicles, which will necessitate the addition of more attack surfaces. Since the CAN bus protocol does not encrypt data, the attackers can use a reverse mechanism to interpret each CAN packet in order to inject malicious messages into in-vehicle networks [7]. This malicious message injection mechanism will cause some abnormal behaviors in the communication traffic, which can be detected by developing an intrusion detection system. The threat of cyber-attacks in the automotive industry and the securing of communication protocols have gotten a lot of attention in recent years. However, due to the complexity of in-vehicle embedded systems and the reality of a real-time experiment with a limited processing unit and memory resources, it is impractical and nearly impossible to apply the standard measures to build a potential IDS for vehicular networks. Therefore, a different mechanism is required to detect normal and abnormal characteristics in a vehicular network. In this manuscript, we propose a deep transfer learning-based intrusion detection model that can efficiently classify the normality and abnormality of a communication traffic and allows the immediate detection of anomalies in the CAN bus protocol. The key contributions of this paper are narrated as follows:

- In this work, a deep transfer learning-based *LeCun Network (LeNet)* model has been proposed for effective intrusion detection in-vehicle network CAN bus protocol. The proposed model enabled to develop effective models that speed up the training process and improve the performance of the deep learning model.
- The experiments have been conducted using an in-vehicle real-time dataset generated from heterogeneous sources that include three types of malicious messages. We have made observations on this practical data to identify the best features in the context of supervised learning for effective intrusion detection.
- In-depth architectural and statistical analyses have been conducted considering traditional machine learning, deep learning, and deep transfer learning algorithms.

Extensive analysis and performance evaluation show that the proposed deep transfer learning-based LeNet model outperforms other approaches.

The rest of this research paper is organized as follows. First of all, Section 2 discusses the background of CAN bus protocol security vulnerabilities and introduces the related work done with in-vehicle networks. Additionally, we present the problem statement and the solution methodology also include the proposed architecture in Section 3. Furthermore, Section 4 discusses a detailed description of the dataset as well as an overview of selected models. Evaluation and experimental results analysis of these methods are discussed in Section 5. Finally, Section 6 includes the summary and feasible future directions for this research.

## 2. Background and Related Work

While investigating the most recent relevant works in this field, we discovered that several of them share a common motivation in different ways. In terms of security features, various studies show the CAN bus's vulnerabilities and weaknesses [8]. The following subsections illustrate those works before we demonstrate our proposed methodology.

### 2.1. Background of CAN and Security Vulnerabilities

During the development of the CAN bus protocol, vehicles were considered as isolated objects that did not have a connection with the outside environment [9]. By design, the CAN bus protocol is plagued with various security issues because of the lack of encryption and authentication requirements [10]. Therefore, any malicious or hijacked node can cause disastrous accidents and serious financial loss. For instance, hackers can affect an ECU by injecting malicious messages and various attacks due to the lack of an efficient message authentication method. The CAN node is the combination of a CAN controller and a CAN transceiver that transmits and receives messages but not simultaneously [11]. The architecture of a CAN bus node is shown by Figure 1. The data frame, remote frame, error frame, and overload frame are four different types of frames that have been used in the CAN bus [9,12]. The data frame is used to transmit actual data from a transmitter to receivers (other nodes). A node requests a specific message with a specific identifier using the remote frame. If any of the nodes on the bus detects an error, it will send an error frame. The overload frame adds a delay between the data and remote frames.

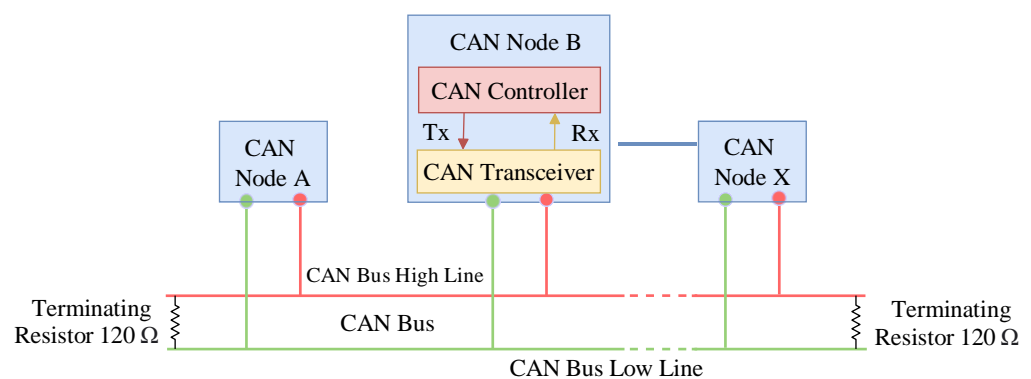

**Figure 1.** The standard CAN bus node architecture.

The standard CAN message frame format is the composition of header, trailer, and protected data payload field that can be up to 64 bits long. The header field is the combination of 1-bit Start of Frame (SOF), 12-bits arbitration field, and 6-bits control field. Furthermore, the arbitration field divided into an 11-bits identifier and 1-bit Remote Transmission Request (RTR) field. The identifier field represents the message priority. It also consists of Identifier Extension (IDE), Reserved, Data Length Code (DLC), Cyclical Redundancy Check (CRC), Delimiter, Acknowledge (ACK), End of Frame (EOF), and Inter Frame Space (IFS) fields. Both sides of the message frame end with a bus idle field. Figure 2 shows the standard CAN message frame format [9].

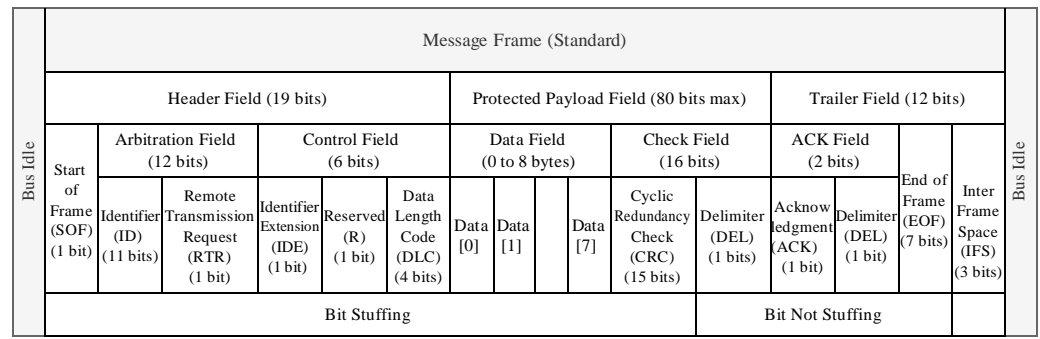

**Figure 2.** The standard CAN bus message frame format.

Nowadays, with the progression of data-mining techniques, many researchers have addressed this type of attack and are able to detect and ignore any abnormal traffic activities in CAN networks [13–17]. Recently, modern vehicles are not considered only a closed-loop system; instead, they also communicate with the outside world via various intelligent systems. As a result, hackers or attackers can use various internal and external interfaces to inject malicious messages into CAN traffic. Despite the tendency of injecting malicious messages, ECUs can be re-programmed remotely by embracing over-the-air (OTA) updates, which may provide more comfort and advantages to the vehicle owner [15]. However, these mechanisms have also initiated more remote attacks, which can assist attackers or hackers in compromising the ECUs by sending malicious messages.

*2.2. TML and DL Based IDSs for IVN*

In the existing automotive applications, the TML and DL-based IDSs have obvious advantages in detecting various malicious messages [18–20]. Bozdal et al. [21] and Lokman et al. [22] reviewed the security threats and challenges of the automotive CAN bus system, and discuss some potential security solutions. In 2016, Kang et al. [23] proposed a Deep Neural Network (DNN)-based IDS for IVN security. Using the unsupervised pre-training model of Deep Belief Networks (DBN), the selected parameters of the DNN model were trained with probability-based feature vectors extracted from the IVN packets, followed by the traditional stochastic gradient descent technique. The results of the experiment showed that the model can provide a real-time response to malicious messages, with a detection ratio over 95% on average in the CAN bus. The robustness of the model is high, but detection coverage is not defined. Loukas et al. [24] proposed cloud-based cyber-physical IDS for IVN using the Deep Learning (DL) model to detect Denial-of-Service (DoS), command injection, and malware (Net) attacks. The model had a validation accuracy score of overall 86.9%, which motivates additional research into this field to improve the detection rate, particularly for these attacks. Seo et al. [25] introduced an IDS to identify DoS, Fuzzy, RPM, and Gear attacks in the CAN bus network traffic by applying Generative Adversarial Network (GAN). This DL-based model can detect unknown malicious messages using only normal CAN data for training. The results of various simulations show that each of the four attacks was detected with a high accuracy score of over 95%, demonstrating the robustness of the model.

Lokman et al. [26] developed an IDS for an in-vehicle network using an unsupervised DL-based model, known as Deep Contractive Auto-encoders (DCAEs). The DCAE model outperformed other regularized auto-encoder variants, with a 91.0% detection rate. As the proposed IDS performance is evaluated within a simulated network, further evaluation is necessary to validate the efficiency against a larger array of various cyber-attacks. Zhang et al. [27] in 2019 proposed a DL-based IDSs for in-vehicle security to detect only spoofing and replay attacks. The results were evaluated in a simulated environment, which is hopeful as they can effectively detect only spoofing and replay attacks. The proposed model is capable of adapting to new attacks. The detection accuracy of this model varies between 97.0% and 98.0% when it faces unknown attack types. Zhu et al. [28] proposed a DL-based method to speed up intrusion detection using the LSTM model. They executed the spoofing, replay, and flooding attacks in the CAN network. The authors proposed using a mobile edge-assisted multi-task LSTM model because the computation time with LSTM is so high. The model had an accuracy score of over 80% and a detection latency of 0.61 ms. Avatefipour et al. [29] proposed a new effective IDS based on a modified one-class SVM in the CAN traffic by deploying three attacks (e.g., DoS, fuzzing, and spoofing attacks). The experimental result shows that the proposed model has a high accuracy score of over 90%, demonstrating the robustness of the model. In order to prove the efficiency of the this model, they applied it to other recent popular public datasets in the scope of CAN bus traffic intrusion detection. Xiao et al. [30] proposed a lightweight ML algorithm based on RNN for IDS on the CAN bus network. The experimental evaluation using appropriate hyper-parameters demonstrated that the proposed model had good performance metrics, compared to LSTM and GAN models.

Al-Saud et al. [31] proposed an IDS model based on an improved SVM model for the CAN bus network. The experimental results on the real dataset reveal the good performance metrics and high robustness of the model against only DoS attacks in electric vehicles. Lin et al. [32] proposed a DL-based intrusion detection system for CAN networks to detect DoS, fuzzing, and impersonation attacks particularly. The model is trained with a deep denoising auto-encoder during the training phase, which includes a feature extraction mechanism. Their work had a low detection rate when compared to other ML algorithms. Yang et al. [33] proposed an IDS using a recurrent neural network with long short-term memory (RNN-LSTM). The selected model had a higher validation accuracy score especially for detecting only spoofing attacks in the CAN network traffic, which motivates additional research into this field to detect other cyber-attacks. A Long Short-Term Memory (LSTM) NN-based IDS was proposed by Hossain et al. [34]. The proposed IDS is capable of detecting various attacks on the CAN bus network, such as DoS, fuzzing, and spoofing attacks. Recently, Song et al. [35] proposed an IDS based on a Deep Convolutional Neural Network (DCNN) model called Inception-ResNet to detect various attacks (e.g., DoS, fuzzing, gear, and RPM attacks) to test in a real-time in-vehicle system. The authors also investigated the sequence of messages for intrusion detection. There are two steps to the proposed model. The first is a training step and the last one is a detection step. In the first step, the CNN classifier is trained and, in the last step, real CAN data frames are passed through this proposed model to classify whether they are normal or attack messages. In comparison to previous work, the proposed model had an over 80% detection rate and a low error rate but has high computational cost and memory consumption. However, further analyses are necessary to investigate the performance on new complex types of cyber-attacks in various categories.

The existing IDSs have categorized according to *detection algorithm*, *detection accuracy*, *robustness*, and *detection coverage*. Here, robustness is defined as the ability of the IDS to detect attacks in the CAN bus network. A summary of all the existing IDSs for in-vehicle network is given in Table 1. Several existing IDSs have used data from different small in-vehicle networks, which can not be implemented in a realistic environment. Moreover, existing IDSs concentrated on detecting whether specific cyber-attacks have occurred, but most of them did not classify the type of attack. This limitation of previous approaches is a significant feature for further investigation for in-vehicle security.

**Table 1.** Overview of recent research on IDSs for in-vehicle networks.

| Ref. | Algorithm | Accuracy | Robustness | Detection Coverage |
|------|-----------|----------|------------|--------------------|
| [23] | DL | >95% | High | N/A |
| [24] | DL | >85% | Medium | DoS, Command Injection, Malware |
| [25] | GAN | >95% | High | DoS, Fuzzing, RPM, Gear attacks |
| [26] | DCAE | >90% | Medium | DoS, Fuzzing, Impersonation |
| [27] | DL | >95% | High | Spoofing, Replay |
| [28] | LSTM | >80% | Medium | Spoofing, Replay, Flooding |
| [29] | ML | >90% | High | DoS, Fuzzing, Spoofing |
| [30] | RNN | >95% | High | DoS, Fuzzing, Impersonation |
| [31] | ML | >90% | Medium | DoS |
| [32] | DL | >80% | N/A | DoS, Fuzzing, Impersonation |
| [33] | RNN-LSTM | >95% | High | Spoofing |
| [34] | NN-LSTM | >90% | N/A | DoS, Fuzzing, Spoofing |
| [35] | DCNN | >80% | Medium | DoS, Fuzzing, RPM, Gear attacks |
| [36] | DTL | >90% | High | Impersonation, ARP, Flooding |

N/A means "Not Applicable".

*2.3. DTL Based IDSs for IVN*

Deep transfer learning (DTL) is a solution that can reuse previous trained-model knowledge and outperform other TML and DL models in terms of intrusion detection [36,37]. Zadrozny et al. [38] proposed a model for intrusion detection that performs better in both labeled and unlabeled data. Another type of transfer learning model called TrAdaBoost was proposed by Dai et al. [39]. This model allows knowledge from the old trained data to be efficiently transferred to the new validation data, resulting in a more efficient classification model. Additionally, Raina et al. [40] also proposed a transfer learning model that builds an informative Bayesian from prior knowledge before validating a new task. Furthermore, Gou et al. [41] proposed a novel transfer learning model for IDS especially to detect the different types of cyber-attacks. The proposed model shows that the detection accuracy of the different types of cyber-attacks has been comprehensively improved than others. Li et al. [36] proposed a transfer learning approach for intrusion detection of different types of attacks on the Internet of Vehicles (IoV). The experimental results show that, when compared to existing TML and DL methods, this model significantly improved detection accuracy by at least 23%. Xu et al. [42] recently proposed an IDS based on DL and transfer learning. To improve the model's efficiency and adaptability, transfer learning is implemented here. The experimental analysis shows that the proposed model outperforms the mainstream TML and DL methods in terms of efficiency and robustness, and it can detect and classify new cyber-attacks more effectively. The deep-computational-intelligence system has recently been applied in transfer-learning to optimize the performance of existing transfer-learning models [37]. As a result, current transfer learning solutions for intrusion detection still need to be updated [36]. A new-generation labeled dataset of an in-vehicle network proposed by Kang et al. [43], which is more suitable for applying transfer learning models because, for time series classification, deep transfer learning approach shows the better performances than other TML or DL models [44–46]. This paper has improved the existing transfer learning model for detecting various complex types of cyber-attacks in CAN bus protocol.

**3. Proposed Solution**

The proposed deep transfer learning-based *LeCun network (P-LeNet)* approach is presented in this subsection. Following that, we have thoroughly explained the problem statement, solution formulation, the structure of the proposed *P-LeNet* based intrusion detection model, and how we adapted it for deep transfer learning.

## 3.1. Problem Statement

Automobile manufacturers are working to develop fully autonomous vehicles, which will necessitate the addition of more attack surfaces. Since the CAN bus protocol does not encrypt data, the attackers can use a reverse mechanism to interpret each CAN packet in order to inject various malicious messages into the in-vehicle network. This malicious message injection mechanism will cause abnormal behaviors in the communication traffic, which can be detected by developing an intrusion detection system. Three types of attacks (e.g., *flooding*, *fuzzing*, and *spoofing*) have been considered due to their severely impaired characteristics, the intensity of an attack, and the degree of damage among in-vehicle functions. The three most common attack scenarios against an in-vehicle network are shown in Figure 3. By maintaining an influential situation on the CAN bus, the *flooding* attack allows an ECU node to hold many of the resources allocated to the CAN bus. This attack disrupts normal driving and limits the communication between ECU nodes by sending high frequency and high priority messages (e.g., 0×000). Figure 3a shows a scenario of a *flooding* attack on CAN networks. In the *fuzzy* attack, a malicious ECU from IVN transmits random frames with spoofed CAN IDs with arbitrary data values, which caused the vehicle function to be unavailable (e.g., 0×4CC, 0×7C6). Due to the limited number of valid CAN frames streaming over the bus, this type of attack is easy to implement and does not necessitate reverse engineering. The *fuzzy* attack scenario against an IVN is shown in Figure 3b. *Spoofing* is a type of attack in which a malicious node transmits messages to the receiver with a fake ID (e.g., 0×2B0, 0×130) that appears identical to that of an original node. As a result, the receiver node considers that the message is from an original node. It is tough to distinguish between malicious and original messages because there is no message authentication mechanism on the CAN bus. Figure 3c shows a scenario for a *spoofing* attack on a CAN network.

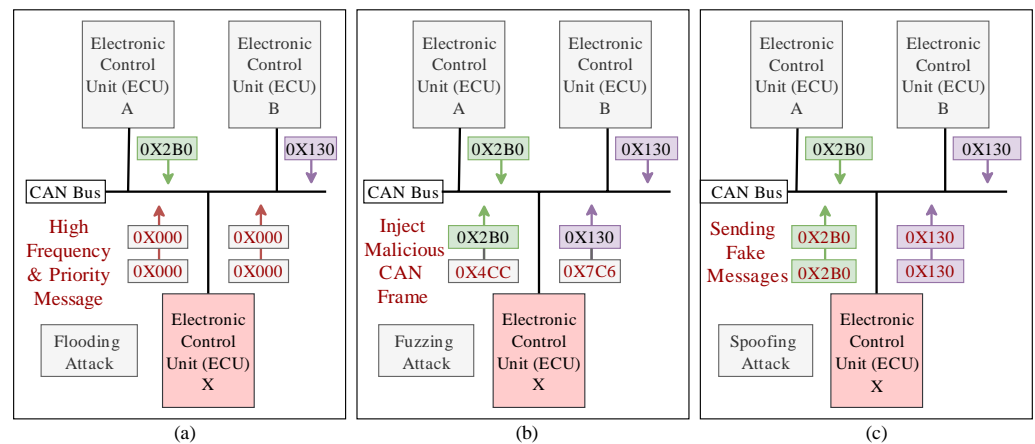

**Figure 3.** (**a**) Flooding attack scenario against an IVN; (**b**) Fuzzing attack scenario against an IVN; (**c**) Spoofing attack scenario against an IVN.

However, due to the complexity of in-vehicle embedded systems and the realities of a real-time experiment with limited processing and memory resources, applying standard measures to build a potential IDS for vehicular networks is impractical and nearly impossible. As a result, detecting normal and abnormal characteristics in a vehicular network requires a different mechanism. The next subsection discusses the solution formation and the details' architecture of our proposed intrusion detection model that can efficiently classify the normality and abnormality of communication traffic and allows the immediate detection of anomalies in the CAN bus protocol. Figure 4 shows the application of our proposed intrusion detection mechanism for vehicular network traffic.

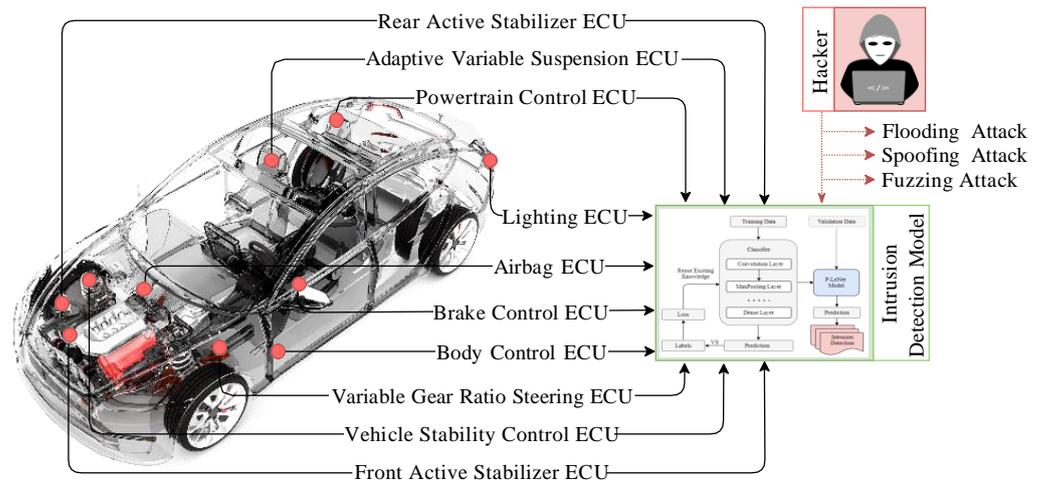

**Figure 4.** Application of intrusion detection model for IVN traffic. The figure illustrates how the proposed IDS can detect the possible attack vectors within an in-vehicle Network. The car image is adopted from [47].

### 3.2. Solution Formulation

Table 2 shows the symbols and descriptions, in which we set the initial model with enough labeled data to build an effective intrusion detection model. The source domain data ($D_s : (A_s, B_s)$) is the combination of ($A_{s1}, B_{s1}$), ($A_{s2}, B_{s2}$), ($A_{s3}, B_{s3}$), ($A_{sn}, B_{sm}$) and the target domain data ($D_t : (A_t, B_t)$) is the combination of ($A_{t1}, B_{t1}$), ($A_{t2}, B_{t2}$), ($A_{t3}, B_{t3}$),..........($A_{tn}, B_{tm}$), in which the class of source domain label data ($B_s$) and target domain label data ($B_t$) is 0 and 1, where the normal and attack scenario is represented by 0 (zero) and 1 (one), respectively.

**Table 2.** Symbols and description.

| Description | Source (s) | Target (t) |
| --- | --- | --- |
| Domain data | $D_s : (A_s, B_s)$ | $D_t : (A_t, B_t)$ |
| Domain feature | $A_s$ | $A_t$ |
| Domain label | $B_s$ | $B_t$ |
| Number of domain data | n | m |

Additionally, both the labels of the source domain ($B_s$) and the target domain ($B_t$) data contain only *normal* and *attack* data, although the attackers in the source and target domains may be different. Although the source domain label ($B_s$) and the target domain label ($B_t$) share the same feature space, they perform differently in specific features. We used the *Maximum Mean Discrepancy* Equation (1) to calculate the difference between the source and target domains [48]:

$$Distance(A_s, A_t) = \left\| \frac{1}{n} \sum_{i=0}^{n} \varphi(A_{s_i}) - \frac{1}{m} \sum_{i=0}^{m} \varphi(A_{t_i}) \right\|^2 \quad (1)$$

The detection model trained by the source domain data ($D_s$) does not have excellent detection accuracy when faced with target domain data ($D_t$), according to the dependency of TML and DL models, and this has been totally proven by the subsequent experiment. The TML and DL models require a large amount of training data. Thus, it is difficult to train an effective IDS model using a small amount of source domain data ($D_s$). As a result, we have proposed a deep transfer learning based *P-LeNet* method to transfer the knowledge contained in source domain data ($D_s$) to the target domain and combine the target domain data ($D_t$) to build an efficient IDS to improve the detection accuracy for any electric vehicular ecosystems.

*3.3. Architecture*

The block diagram of the proposed *P-LeNet* model is shown in Figure 5, which contains two parts: the model training part and the intrusion detection part. After pre-processing the raw data, we have applied it to our proposed model for training. Through subsequent empirical experiments, the most important parameters for the selected model have been determined. We used a randomly selected training dataset to train the proposed *P-LeNet* model and a validation dataset to validate the model. The final IDS model has been selected based on its best prediction performance on the validation dataset.

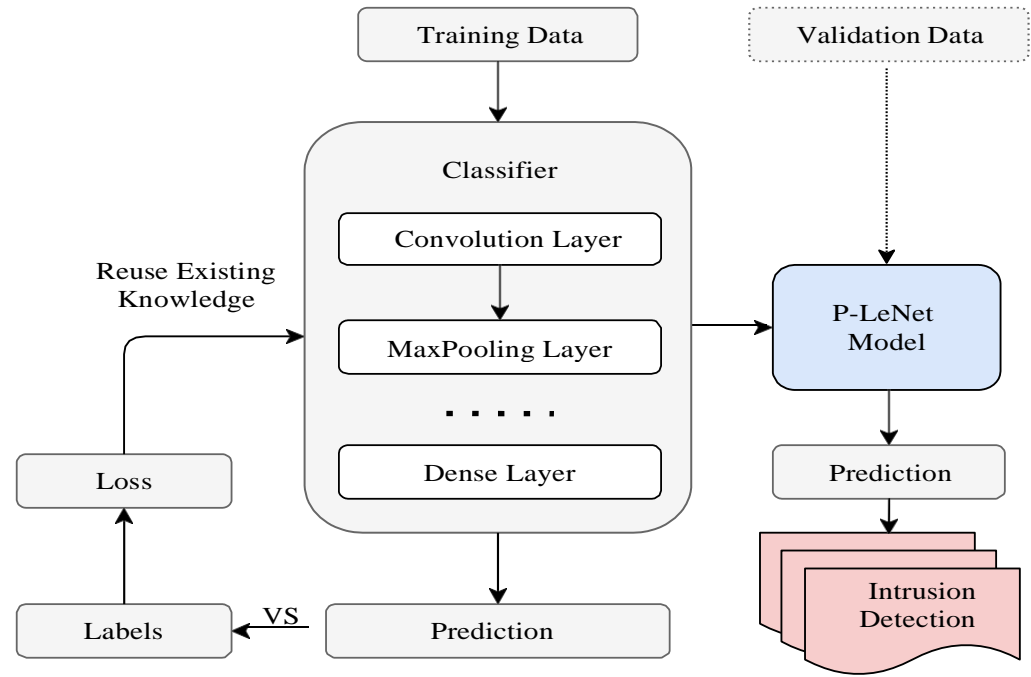

**Figure 5.** The block diagram of the proposed P-LeNet mode.

The proposed *P-LeNet* architecture is made up of seven layers with a total of 12,052 trainable parameters (weights). The layer is the composition of two *convolutional layers*, two *subsampling layers*, one *flatten layer*, one *fully connected layer*, and one *output layer*. Each layer takes the previous layer outputs as inputs for the current layer and performs some nonlinearity's to transform it into a multivariate series whose dimensions are defined by the number of filters in each layer. The structure of the proposed *LeCun Network (P-LeNet)* model is shown in Figure 6. The first layer is the *Input* layer, which is not considered a network layer because it does not learn anything. The input layer is designed to take dataset and pass it on to the following layer. The dataset has a total of four features including the label feature. The four features are *CAN_ID*, *DLC*, *Data_Field*, and *Label*. The one-dimensional convolutional layer *(Conv1D)* is used in the first, and the third layer respectively to transform the dataset. The first *Conv1D* layer produces as output five feature maps, and has a kernel size of 5, and the second *Conv1D* layer produces as output 20 feature maps, and has a kernel size of 5. The *Rectified Linear-Unit (ReLU)* activation function is used in the both convolution layer. The two *Conv1D* layers contain 30 and 520 trainable parameters, respectively. The first *MaxPooling1D* subsampling layer follows the first *Conv1D* layer, and the second *MaxPooling1D* subsampling layer follows the second *Conv1D* layer shown in Figure 6. The two subsampling layers halves the dimension of the feature maps it receives from the previous layer; this is known commonly as downsampling. The two subsampling layers also produce 5 and 20 feature maps, respectively, each one corresponding to the feature maps passed as input from the previous layer. The fifth layer of our proposed model is the *Flatten* layer which converts the pooled feature map to a single column that is passed to the next layer. The next is fully-connected *Dense* layer where total trainable parameter is 10,500. This operation reduces drastically the number of trainable parameters in a deep model while enabling the use of a class activation map which allows an interpretation of the learned features [49]. Finally, the output layer whose number of neurons is equal to the number of classes in the dataset. The *softmax* function is used as the activation function in this layer to predict a probability distribution between normal and attack scenarios.

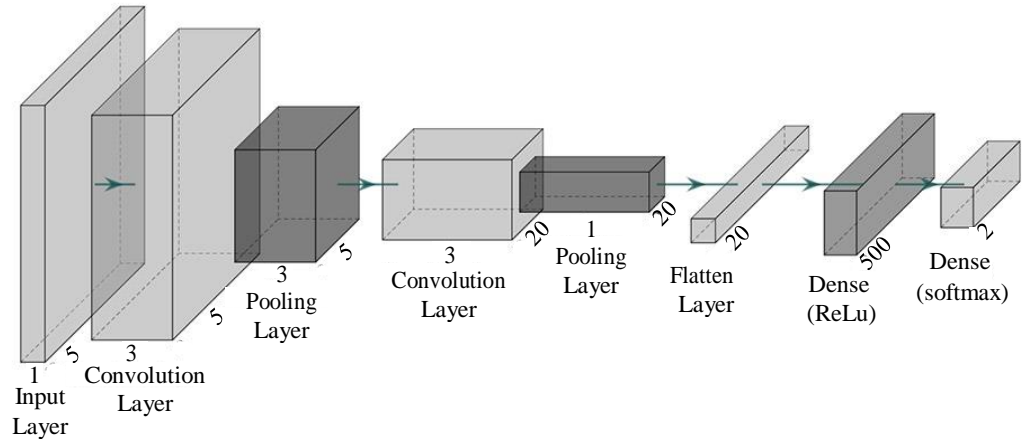

**Figure 6.** The structure of the proposed P-LeNet model.

Furthermore, the *Compile* function enables the actual building of the model we have implemented with some additional characteristics such as the *loss function*, *optimizer*, *learning rate*, and *metrics*. To train the network, we utilize a loss function called *categorical crossentropy*, which calculates the difference between the network's predicted values and the actual values of the training data. The number of changes made to the weights within the network is facilitated by the loss values accompanied by an optimization algorithm *(Adam)*. During training, we have been used the valuation dataset to validate our proposed model after each epoch. The proposed model has achieved a better validation accuracy. However, we have evaluated the trained model on the test dataset for a more explicit verification of the proposed model's performance on an unknown dataset.

## 4. Materials and Methods

In this section, we have thoroughly explained the dataset and the transformation process of the dataset to feed the selected models.

### 4.1. Dataset Description

The dataset has been generated in two different ways. Details of the dataset can be accessed in [43]. The first dataset contained normal driving data without an attack and the second dataset contained abnormal driving data that has been collected during an attack was performed in in-vehicle networks. Each dataset has been combined into one CSV file by a Python script. The class distribution of the combined dataset is shown in Figure 7. The combined dataset has a total of 5 features including the label feature. The five features are *Timestamp*, *CAN_ID*, *DLC*, *Data_Field*, and *Label*. The *Timestamp* feature represents the recorded time in seconds (s). The *CAN_ID* is used to identify the CAN messages in hexadecimal format (e.g., 0xFA5, 0x18F) and assigns its priority. The messages having the lowest *CAN_ID* value represent the highest priority. The *DLC* feature in the control field shows the number of bytes, from 0 to 8, and values change depending on the vehicle categories. The *Data_Filed* feature contains the data to be transferred from one node to another and consists of the data value in a byte that has eight fields in total (e.g., Data[0], Data[5]). Finally, the *Label* feature contains two quantitative values, i.e., 0 and 1, which indicates normal and attack (injected message) scenarios, respectively.

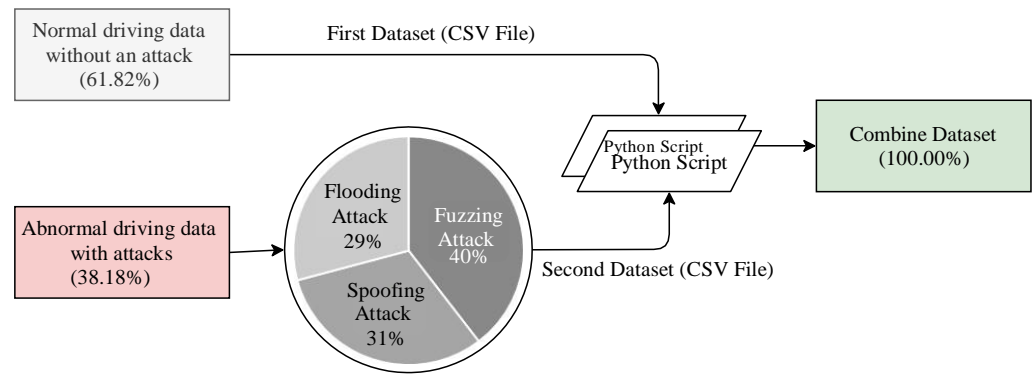

**Figure 7.** Statistics of the dataset.

*4.2. Data Preparation*

The dataset must be cleaned and prepared before applying the selected TML, DL, and DTL methods to achieve optimal performance and improve the learning process. Data preparation generally happens by eliminating unnecessary features, checking for changes in independent features, converting non-numeric features, and removing outliers. Three fundamental steps are applied during the data preparation process. The first step is data cleaning, the second is data integration, and the final step is data transformation.

4.2.1. Data Cleaning

This dataset is very sensitive to missing and noisy data because of its large size. There are a total of 1,270,310 instances in the dataset including noisy and inconsistent data. In this subsection, we have discussed the essential steps in prepossessing of data. First of all, we have applied various techniques to remove noise and clean inconsistencies data from the dataset, for example, *Rosner's Test* for outliers checking, and the *Predictive Mean Matching* method for imputing missing values. Then, in order to apply the selected models, we have converted the qualitative values into quantitative values. To cite an example, the *Label* feature in the dataset, which has qualitative values *'Normal'* and *'Attack'*, has been converted into '0' and '1'. These quantitative values have been converted to quantitative values by performing a numerical convolution label-encoding library *numconv*. The *CAN_ID* feature in the dataset, which has hexadecimal values (e.g., 58B, F41), has been converted into decimal values by applying the *hex2dec* function. On the other hand, the *Data_Field* feature in the dataset, which also has hexadecimal values of eight bytes separated by space (e.g., 80 7F 00 73 20 00 0A A1, 14 80 10 80 00 00 0A 73). The space between bytes have been removed by applying *gsub* function and then the hexadecimal values have been converted into decimal values by applying the *Rmpfr* function as most of the data field is over 64 bits (maximum 152 bits). The *Timestamp* feature has been omitted from feature vectors as they may cause overfitting the training data. Furthermore, for some DL and DTL models, the input data shape has been reshaped into three dimensions to feed the models by applying *numpy.reshape* with *swapaxes* and *concatenate* methods.

4.2.2. Data Integration

To improve the accuracy and speed of the training and validation process, the data integration technique helped us by reducing and avoiding redundancies from the resulting dataset. As this dataset originates from two different ways. Thus, it is an essential step to analyze the *redundancy* and *correlation* between the selected features. This analysis has measured how strongly one feature, i.e., *CAN_ID* implies the other, i.e., *Data_Field*. We used cutoff criteria ($p < 0.05$) to find the correlation between different features. The results indicated that the higher the coefficient value, the stronger the relationship between those features [50]. Table 3 shows the correlation between different features. For our analysis, we assessed the correlation between all features by calculating the following *Pearson product-moment coefficient* Equation (2) [51]:

$$r = \frac{\sum_{i=1}^{n}(x_i - \bar{x})(y_i - \bar{y})}{\sqrt{\sum_{i=1}^{n}(x_i - \bar{x})^2 \sum_{i=1}^{n}(y_i - \bar{y})^2}} \qquad (2)$$

where $n$ is the number of tuples, $x_i$ and $y_i$ are the respective values in tuple $i$, and $\bar{x}$ and $\bar{y}$ are the respective mean values of $x$ and $y$.

**Table 3.** Pearson product–moment correlation between different features.

|            | Timestamp              | CAN_ID                 | DLC                    | Data_Field             |
|------------|------------------------|------------------------|------------------------|------------------------|
| **Timestamp** | $0 \times 10^{0}$    | $4.285 \times 10^{-2}$ | $6.525 \times 10^{-6}$ | $1.794 \times 10^{-4}$ |
| **CAN_ID**    | $4.285 \times 10^{-2}$ | $0 \times 10^{0}$    | $1.663 \times 10^{-1}$ | $3.966 \times 10^{-1}$ |
| **DLC**       | $6.525 \times 10^{-6}$ | $1.663 \times 10^{-1}$ | $0 \times 10^{0}$    | $2.707 \times 10^{-1}$ |
| **Data_Field** | $1.794 \times 10^{-4}$ | $3.966 \times 10^{-1}$ | $2.707 \times 10^{-1}$ | $0 \times 10^{0}$    |

Cutoff criteria: $p < 0.05$ (statistically significant) and the values have been rounded to the four decimal places.

### 4.2.3. Data Transformation

We have taken this step to achieve more efficient results and to better understand the patterns. Some features are higher than others, leading to wrong performance, though some models may be preferred for larger functional values. We have performed these strategies to re-scale the selected feature values within a range between [0.0, 1.0] without changing the characteristics of original data [52,53]. As shown in the following Equation (3), a technique called *minimum–maximum normalization* has been used to re-scale the selected feature values within the range:

$$N_v = \frac{X - X_{min}}{X_{max} - X_{min}} \qquad (3)$$

where $N_v$ is the output normalized values, $X$ is an original value and $X_{max}$, and $X_{min}$ is the maximum and minimum values of the feature, respectively.

### 4.3. Training Process

As mentioned in the previous subsection, a Python script combined the two datasets into a single dataset that included both the training and test data. First of all, we have used the *scikit_learn* library's *train_test_split* method to split the combined dataset into the training (80%) and test (20%) datasets. In the raw dataset, the total number of data are 1,270,310. After removing the noisy and inconsistent data, we got a total of 1,257,303 data where the number of training data are 1,005,843 (80%) and the testing data are 251,460 (20%). The training dataset has been used to train the selected models, and the test dataset has been used to further assess the trained classifier. Furthermore, we split again the training data 1,005,843 (80% of the total data) into the new training data 804,674 (80%) for training the selected model and validation data 201,169 (20%) for hyperparameters' optimization. The percentages of 80% for the training dataset and 20% for the test dataset have been chosen as suggested in [54]. To avoid the over-fitting problem, this splitting ratio has been considered as the best ratio between the training and the test dataset [55]. We have used the value of the *random_state* parameter as true, which decided the splitting of dataset into the training and the test dataset randomly [56]. Finally, various performance indicators have used to evaluate the overall performance of the selected models, which have been discussed in the *Results* section. The steps involved to evaluate the performance of all of the selected models are summarized in Figure 8.

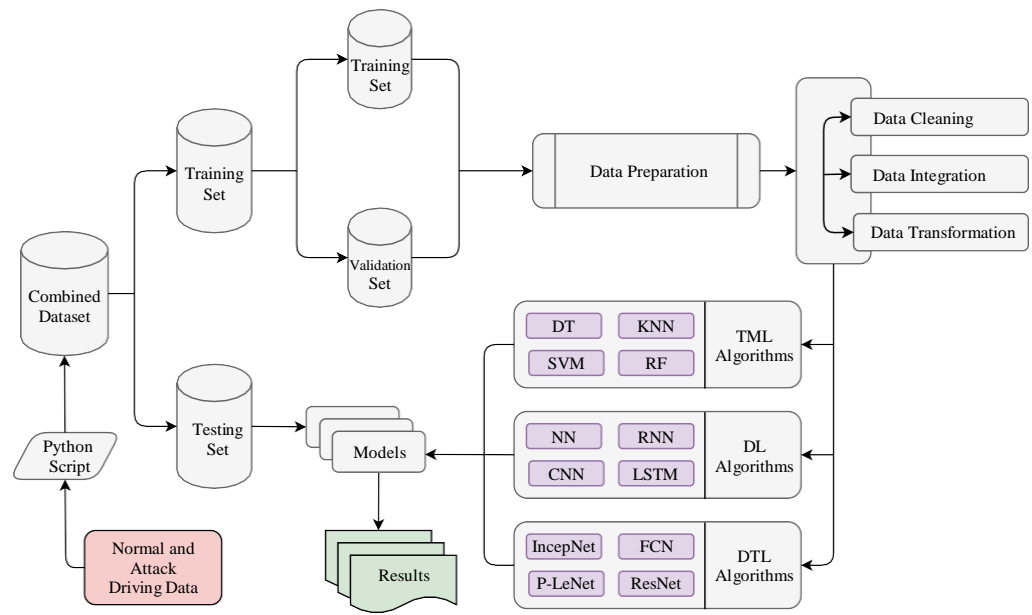

**Figure 8.** Evaluation process on the datasets with the selected models.

Several supervised TML algorithms have been applied to evaluate their performance for intrusion–detection purposes. The TML algorithms have been chosen based on their extensively used in the security-domain as they have already shown good performance on these scenarios [57]. We have predominantly applied Decision Tree (DT), Random Forest (RF), Support Vector Machine (SVM), and K-Nearest Neighbor (KNN) algorithms for classification analysis. The DT algorithm is the most popular model that is used for the IDS domain, which is showed by the authors in [57,58]. The effectiveness of the RF models in the IDS domain has been shown in a survey conducted by Yang et al. [59]. The SVM algorithm has been considered as it has low computation overheads [60]. Finally, the KNN algorithm has been selected as it achieves good performance in dealing with different sensor data [57].

In recent years, DL algorithms have advanced significantly and some of the variants of DL algorithms have been successfully applied to solve classification tasks related to intrusion detection [61]. Therefore, we have considered simple Neural Network (NN), Recurrent Neural Network (RNN), Convolutional Neural Network (CNN), and Long Short-Term Memory (LSTM) algorithms because of their optimal performance. In the LSTM algorithm, we have used three hidden layers, and the union of hidden layers are 128, 100, and 64, respectively. The *tanh* is the hidden layer activation function and *Adam* is used as an optimizer. In addition, *sigmoid* is used as a network output activation function, and *categorical_crossentropy* is used as a loss function. Furthermore, both NN and CNN algorithms have used the same optimizer and activation function but a different type of hidden layer activation function. Particularly for the RNN model, the *softmax* is used as the network output activation function, and *categorical_crossentropy* is used as the loss function. On the other hand, for the CNN algorithm, the number of hidden layers is four, and *binary_crossentropy* is used as the loss function. The dropout layer has been added after each layer to prevent model overfitting as RNN and LSTM generally have the problem of overfitting [62]. We have evaluated the selected DL models with a wide range of tested hyperparameters. We have obtained the optimal performance when we used these combinations of tested hyperparameters. Table 4 shows the hyper-parameter list of all the selected DL algorithms.

**Table 4.** The hyper-parameters of the selected DL models.

| Parameters | NN | RNN | CNN | LSTM |
|---|---|---|---|---|
| Number of hidden Layers | 2 | 3 | 4 | 3 |
| Units in hidden layers | 68, 68 | 64, 64, 64 | 32, 64, 256, 128 | 128, 100, 64 |
| Batch size | 64 | 64 | 64 | 16 |
| Hidden layer activation | relu | relu | relu | tanh |
| Output activation function | sigmoid | softmax | sigmoid | sigmoid |
| Dropout | N/A | 0.1 | N/A | 0.2 |
| Optimizer | Adam | Adam | Adam | Adam |

In addition, we would like to emphasize some DTL models that give better performance than others. We have considered four DTL models and the selected model are Fully Convolutional Networks (FCN), Inception Network (IncepNet), Residual Neural Network (ResNet), and our proposed LeCun Network (LeNet). For all the selected models, the hyper-parameters—batch size, hidden layer activation function, output layer activation function, loss function, and the optimizer are *64*, *ReLu*, *softmax*, *categorical_crossentropy*, and *Adam*, respectively. Furthermore, FCN, IncepNet, and ResNet models have used the same number of hidden layers, but the units in the hidden layer are different. Particularly for the P-LeNet model, the number of the hidden layers is 2, and the units in the hidden layer are 5, 20. We have evaluated the selected DTL models with a wide range of tested hyperparameters. We have obtained the optimal performance when we used these combinations of tested hyperparameters. On the other hand, we have used the Adam optimizer for all the models because it combines the best properties of the *AdaGrad* and *RMSProp* algorithms to provide an optimization algorithm [63]. Furthermore, particularly for this analysis, the *Adam* optimizer has shown the lowest training loss and validation loss among other optimizers. Table 5 shows the hyper-parameters list of all the selected DTL algorithms.

**Table 5.** The hyper-parameters of the selected DTL models.

| Parameters | FCN | IncepNet | ResNet | P-LeNet |
|---|---|---|---|---|
| Number of hidden Layers | 3 | 3 | 3 | 2 |
| Units in hidden layers | 128, 256, 128 | 32, 64, 32 | 128, 256, 128 | 5, 20 |
| Batch size | 64 | 64 | 64 | 64 |
| Hidden layer activation | relu | linear | relu | relu |
| Output activation function | softmax | softmax | softmax | softmax |
| Dropout | N/A | N/A | 0.1 | N/A |
| Optimizer | Adam | Adam | Adam | Adam |

## 5. Results

This section discusses the overall performance of the selected models, starting with an analysis of TML metrics and concluding by explaining the effectiveness of the DL and DTL models.

### 5.1. Experimental Evaluation Indicators

Various measurement indicators (e.g., accuracy, precision, F1-score) are used to illustrate the results where the obtained accuracy shows the overall effectiveness of the proposed model. We evaluated the performance of all the selected models using the following four terms:

- *True-positive (TP)* refers to the number of actual attack instances that are correctly detected as attack.
- *True-negative (TN)* is the number of normal instances that are correctly detected as normal.

- *False-positive (FP)* is the number of normal instances that are incorrectly detected as attack.
- *False-negative (FN)* refers to the number of actual attack instances that are incorrectly detected as normal.

Accuracy is the closeness of the measurements to a specific value, which demonstrates the efficiency of the classifier to determine the total instances. Clearly, a higher accuracy means better classification results. The mathematical expression of accuracy can be defined as follows:

$$Accuracy = \frac{TP + TN}{TP + TN + FP + FN} \quad (4)$$

The fraction of the true positive instances in the positive case determined by the classifier is represented by precision, which demonstrates the closeness of the measurements to each other. Precision can be represented by the following equation:

$$Precision = \frac{TP}{TP + FP} \quad (5)$$

The proportion of relevant instances (positive cases) that are correctly judged to the total positive case is referred to as recall. Recall can be defined as follows:

$$Recall = \frac{TP}{TP + FN} \quad (6)$$

The following F1-score computes the harmonic mean of precision and recall, respectively. F1-score can range from 1.0 to 0.0, with 1.0 indicating perfect precision and recall:

$$F1Score = 2 \times \frac{Precision \times Recall}{Precision + Recall} \quad (7)$$

As one of the significant indicators, the ROC AUC determines areas where the proposed model is classified better within normal and attack scenarios. Measuring ROC AUC requires diagnostic accuracy, which depends on the sensitivity, i.e., true positive rate (TPR), and the specificity, i.e., true negative rate (TNR). As demonstrated by the following equations, TPR, often called recall:

$$Sensitivity(TPR) = \frac{TP}{TP + FN} \quad (8)$$

$$Specificity(TNR) = \frac{TN}{TN + FP} \quad (9)$$

In this analysis, we incorporated a wide range of analyses' scenarios with varying parameters. To this end, we conduct experiments considering different numbers of the hidden layers, units in the hidden layers, numbers of the epoch, a range of hidden layer activation functions, output layer activation functions, loss functions, and different optimizers. These critical parameters significantly affect the calculation of the performance metrics for DL and DTL algorithms. Finally, we show the performance comparisons between TML, DL, and DTL models where the performance of the proposed model has a meaningful impact to indicate the predicted label correctly.

### 5.2. TML Models Analysis

We start with the traditional machine learning (TML) algorithms because these state-of-the-art algorithms provide the optimal performance and take the least amount of time to run. The performance of the selected TML algorithms has been quantitatively evaluated using the fundamental evaluation indicators. Figure 9 shows the performance metrics of the selected TML algorithms. The DT algorithm shows the optimal performance among all considered TML algorithms with an accuracy score of 0.9532, precision score of 0.9463, recall score of 0.9558, F1-score of 0.9608, and ROC AUC score of 0.8408. As shown in Figure 9, the red quadrangle represents the accuracy score for all of the selected TML algorithms and the best one highlighted in blue color text. On the other hand, the KNN algorithm shows the lowest accuracy score of 0.9248 and a precision score of 0.9141. We used an accuracy plot to determine the proper K value for the KNN algorithm, and the highest accuracy was obtained when K = 12. The RF algorithm achieves the second-best performance with an accuracy score of 0.9448 and a precision score of 0.9448. The SVM algorithm achieves an accuracy score of almost 0.9440 and a precision score of 0.8840 by using C = 1.2, epsilon = 0.1, and *Gaussian RBF Kernel* particularly, which is the third-best performance. However, RF and SVM algorithms have an overall accuracy score of almost 0.9444, but the SVM algorithm has the lowest precision score of 0.8840 when compared to the other selected TML algorithms. The lowest precision score of the SVM algorithm indicates that most of the predicted labels are incorrect. In contrast, the precision score of the RF and DT algorithms is around 0.9455, indicating that the majority of predicted labels are correctly classified. Taking into account all aspects of performance indicators, we conclude that the DT algorithm outperforms the other selected TML algorithms. This optimal performance indicates that most of the predicted labels are correctly classified between normal and attack scenarios.

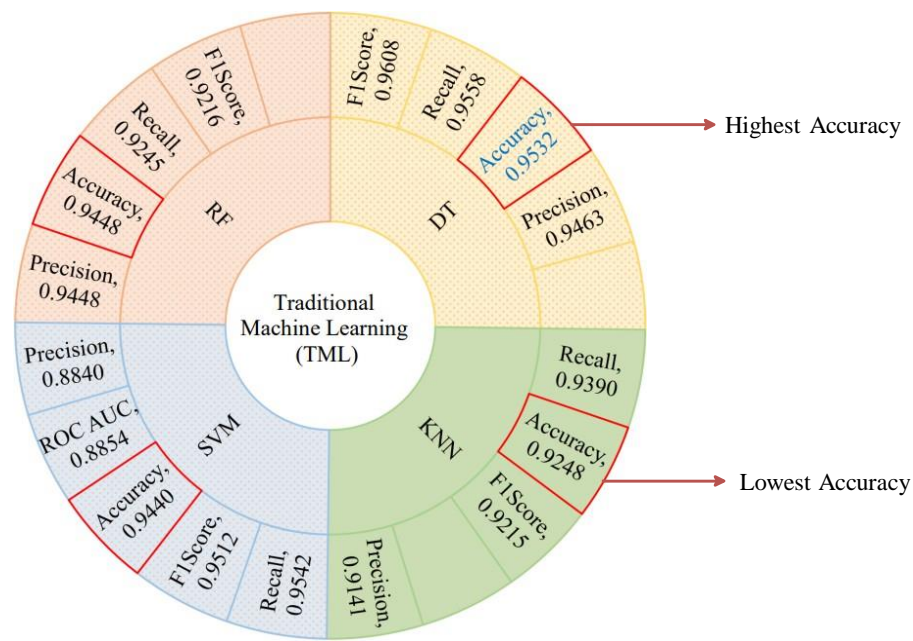

**Figure 9.** TML algorithms' performance metrics visualization.

*5.3. DL Models Analysis*

In recent years, DL models have advanced significantly, and several variations of these models have been successfully applied to solve classification tasks related to intrusion detection [61]. Therefore, in this subsection, we consider DL models because of their optimal performance. We have considered LSTM, NN, CNN, and RNN models. The LSTM model shows the highest accuracy score of 0.9762, precision score of 0.9808, recall score of 0.9392, F1-score of 0.8884, and ROC AUC score of 0.9288. Figure 10 shows the performance metrics comparison of the selected DL models. We used three hidden layers for the LSTM model, with units of hidden layers is 128, 100, and 64. The *tanh* is the hidden layer activation function and *Adam* is used as optimizer. In addition, *sigmoid* is used as a network output activation function, and *categorical_crossentropy* is used as a loss function. On the other hand, the NN model has the lowest accuracy score of 0.9563. The performance of the RNN and the CNN models outperform most of the TML models, where these models show an accuracy score of 0.9640 and 0.9590, respectively.

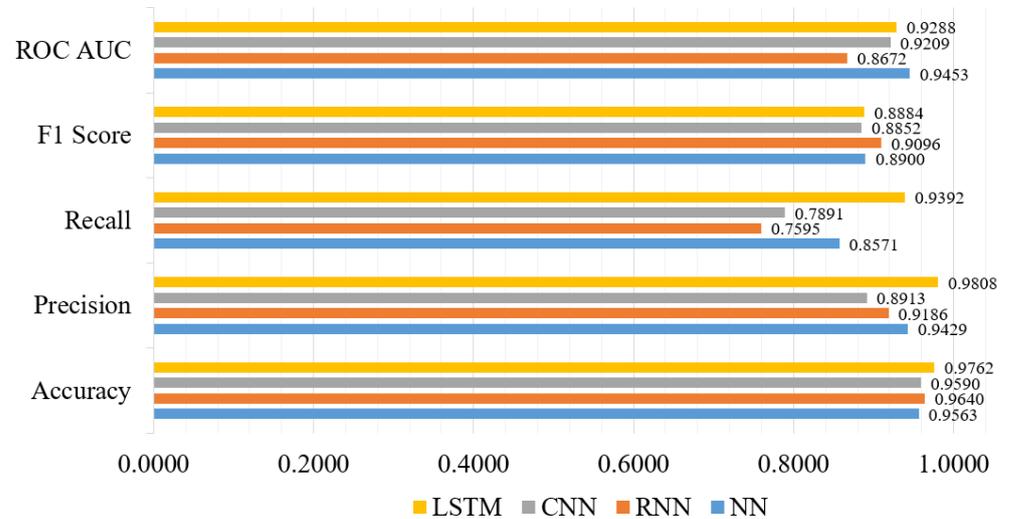

**Figure 10.** DL models' performance metrics visualization.

Furthermore, both the NN and CNN models used the same optimizer and activation function, but different types of hidden layer activation functions. Particularly for the RNN model, the *softmax* is used as the network output activation function, and *categorical_crossentropy* is used as the loss function. On the other hand, for the CNN model, the number of hidden layers is four, and *binary_crossentropy* is used as the loss function. Finally, taking into account all of the significant parameters of the DL models, the LSTM model demonstrated the best performance, indicating that the majority of the predicted labels are correctly classified.

*5.4. DTL Models Analysis*

We analyzed the results of TML and DL models to achieve optimal performance. In this subsection, we highlighted four selected DTL models that perform better than others. We considered the number of hidden layers, units in the hidden layers, output layer activation functions, loss functions, and so on, in addition to the fundamental evaluation criteria. First of all, we considered the quantitative performance of DTL models. Table 6 shows the quantitative performance summary of the DTL models, where the proposed *P-LeNet* model shows an optimal performance compared to the other DTL models with an accuracy score of 0.9810, precision score of 0.9814, recall score of 0.9804, F1-score of 0.9783, and ROC AUC score of 0.9542. In this model, we used two hidden-layers where *relu* is the hidden layer activation function. In addition, *softmax* is used as a network output activation function, and *categorical_crossentropy* is used as a loss function along with *adam* optimizer.

**Table 6.** DTL models' performance comparison metrics.

| Algorithm | Accuracy | Precision | Recall | F1-Score | ROC AUC |
|---|---|---|---|---|---|
| FCN | 0.9786 | 0.9832 | 0.9617 | 0.9488 | 0.9248 |
| IncepNet | 0.9803 | 0.9152 | 0.9265 | 0.9024 | 0.9129 |
| ResNet | 0.9795 | 0.8958 | 0.8845 | 0.9001 | 0.8703 |
| LeNet | 0.9810 | 0.9814 | 0.9804 | 0.9783 | 0.9542 |

The values have been rounded to the four decimal places.

Figure 11 shows the accuracy score of every single epoch for both the training and testing phase on the above-mentioned DTL models. We have considered 1000 epochs for our analysis because the flattening characteristics of the curve and the training/testing accuracy are not increasing literally between the epoch number 450 to 1000. The proposed *P-LeNet* model has the highest accuracy score in both the training and testing phases. For a better understanding of our proposed *P-LeNet* model, Figure 12 shows the trend of the accuracy score in both phases. The proposed model's accuracy increases rapidly in epoch number 10 and gradually rises to a point close to 0.9809 at epoch number 400. However, as shown in Figure 12, which remains nearly stable up to the early stopping checkpoint with an accuracy score of 0.9810. In contrast, the FCN model has the lowest accuracy score in the epoch number ranges from 1 to 1000. The FCN model's accuracy begins around 0.9545 for the training phase and 0.9422 for the testing phase in epoch number 46, as shown in Figure 11. However, it rises dramatically around 0.9785 in epoch 554 and 0.9783 in epoch 610 for the training and testing phases, respectively. Furthermore, the accuracy score of 0.9786 remains stable in epochs 555 to 1000 during the training phase. On the other hand, for the testing phase, the accuracy score of 0.9783 remains stable between epoch numbers 611 and 1000.

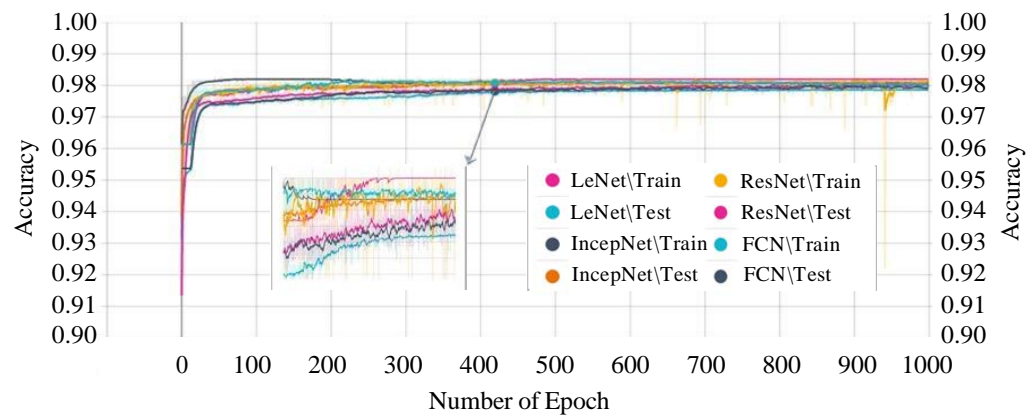

**Figure 11.** DTL models' training and testing accuracy.

The training and testing accuracy of the IncepNet model remains steady between the epoch number 500 to 1000 as shown in Figure 11. The accuracy of ResNet models starts with a score of 0.9102 for training and 0.9089 for the testing phase. However, as the epoch number increases, this score gradually rises and reaches approximately 0.9745 when the epoch number is 450, and then remains stable between epoch numbers 451 and 1000 for both phases. The remarkable point is that the behavior of the training phase is nearly identical to that of the testing phase. For better understanding, the trend of both phases is shown by zooming in Figure 11.

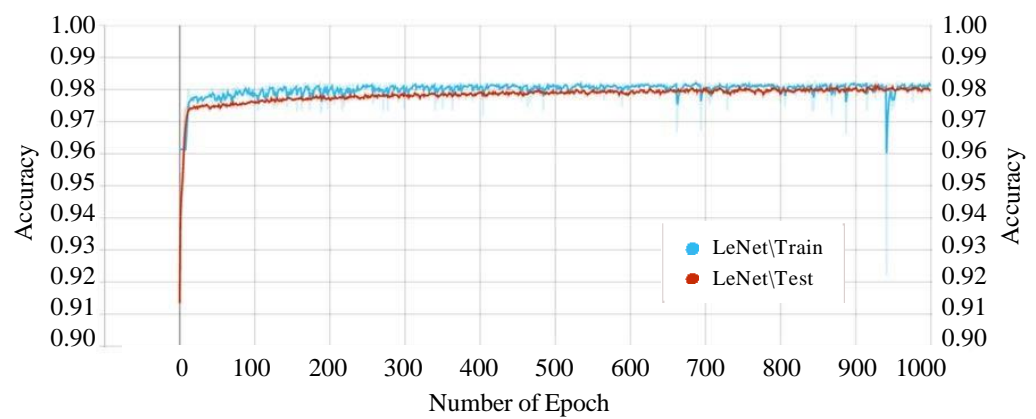

**Figure 12.** Training and testing accuracy of the proposed P-LeNet model.

Next, we analyzed the losses of each model. The training and testing phases' losses of every single epoch are shown in Figure 13. The FCN model shows the highest loss in both the training and testing phase. The highest loss of this model indicates that the model cannot provide a reliable classification between normal and attack scenarios in the CAN networks.

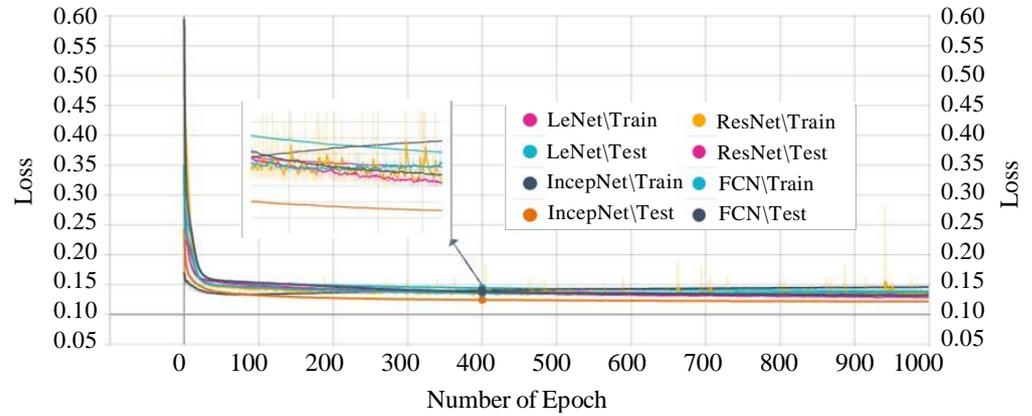

**Figure 13.** DTL models' training and testing loss.

On the other hand, our proposed *P-LeNet* model shows the lowest loss. In detail, the loss of the proposed model starts around 0.50 for the training phase and 0.35 for the testing phase at the beginning as shown in Figure 14. However, it is decreasing to 0.1689 and 0.1750 for the epoch number 10 for the training and testing phase, respectively. The loss score of 0.1459 remains almost constant between epochs 400 and 1000. Furthermore, the losses of the IncepNet and ResNet models are nearly constant during both the training and testing phases, as shown in Figure 13. The loss of these models is almost 0.5506 at the beginning, which declines gradually to approximately 0.1409 for the epoch number 350 and remains stable for both of the phases. For better understanding, the trend of both phases is shown by zooming in Figure 13.

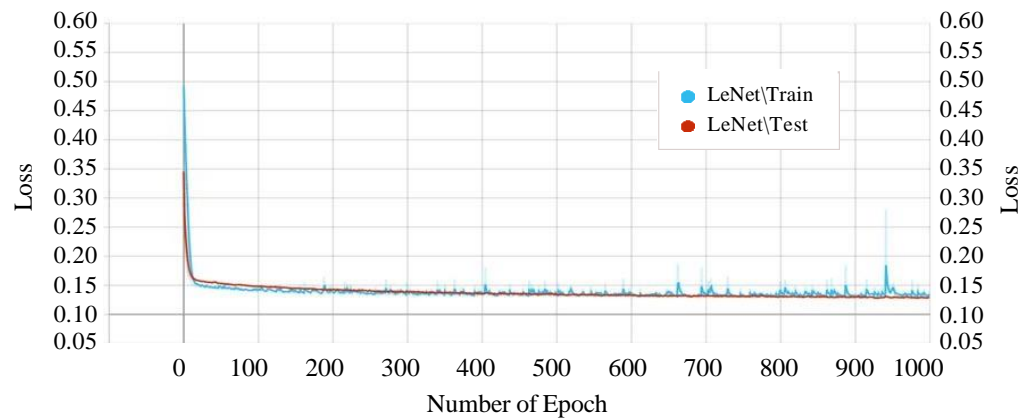

**Figure 14.** Training and testing loss of the proposed P-LeNet model.

### 5.5. Performance Comparison

We have considered a variety of strategies for selecting the important features and then applied the chosen algorithms. We have predominantly applied the TML, DL, and DTL approach in the same dataset. The TML algorithms did not show remarkable performance. However, when we have applied the DL algorithms, most of the algorithms perform better than the TML algorithms. Furthermore, when we have applied the DTL algorithms, all the algorithms perform significantly better in most cases. In such cases, accuracy, precision, recall, F1-score, and ROC AUC are better than other experimental scenarios. Moreover, the proposed *P-LeNet* model has adequate stability, low loss, and better classification accuracy than other DTL approaches. Finally, the proposed model can effectively identify and classify the normal and attack scenarios of in-vehicle networks to correctly manage vehicle communications for vehicle security.

## 6. Conclusions

Automobile manufacturers are working to develop fully autonomous vehicles, which will ensure proper security. In this manuscript, we propose a deep transfer learning-based LeNet model for intrusion detection in electric in-vehicle networks. The proposed detection model has an overall accuracy score of 98.10%. Moreover, the model has precision score 98.14%, recall score 98.04%, F1-score 97.83%, and ROC AUC score by 95.42%, which is a noticeable improvement when compared to the other benchmark ML, DL, and DTL models. These experimental results demonstrated that the proposed *P-LeNet* model efficiently classifies the normality and abnormality and allows the immediate detection of anomalies in the CAN networks. To summarize, it is obvious that the model has proven its potential to efficiently exhibit anomalous data identification to protect the CAN network that can also be extended in other emerging applications within critical infrastructures where automation and secure data processing is the main challenge. In the future, we will try to implement this proposed deep learning model based on decentralized devices or servers. We will also concentrate on improving the performance of the proposed model by optimizing the hyper-parameters.

**Author Contributions:** Conceptualization, S.T.M. and A.A.; methodology, S.T.M. and A.A.; software, S.T.M., K.A., and Z.R.; validation, S.T.M., A.A., Z.R., and K.A.; formal analysis, S.T.M. and A.A. and Z.R.; investigation, S.T.M., A.A., and K.A.; resources, S.T.M., Z.R. and A.A.; data curation, S.T.M.; writing—original draft preparation, S.T.M.; writing—review and editing, A.A., Z.R., and K.A.; supervision, A.A., Z.R., and K.A.; project administration, A.A.; funding acquisition, A.A. All authors have read and agreed to the published version of the manuscript.

**Funding:** This research received no external funding.

**Institutional Review Board Statement:** Not applicable.

**Informed Consent Statement:** Not applicable.

**Data Availability Statement:** The data presented in this manuscript are available on request from the corresponding author.

**Conflicts of Interest:** The authors declare no conflict of interest.

## Abbreviations

The following abbreviations are used in this manuscript:

| | |
|---|---|
| CAN | Controller Area Network |
| ECU | Electronic Control Unit |
| IVN | In-Vehicle Network |
| CAN FD | CAN Flexible Data-Rate |
| LIN | Local Interconnect Network |
| MOST | Media Oriented Systems Transport |
| ML | Machine Learning |
| IDS | Intrusion Detection System |
| TML | Traditional Machine Learning |
| DL | Deep Learning |
| DTL | Deep Transfer Learning |
| DT | Decision Tree |
| RF | Random Forest |
| SVM | Support Vector Machine |
| KNN | K-Nearest Neighbor |
| NN | Neural Network |
| RNN | Recurrent Neural Network |
| CNN | Convolutional Neural Network |
| LSTM | Long Short-Term Memory |
| FCN | Fully Convolutional Networks |
| IncepNet | Inception Network |
| ResNet | Residual Neural Network |
| LeNet | LeCun Network |
| ReLu | Rectified Linear-Unit |
| DLC | Data Length Code |
| TNR | True Negative Rate |
| TPR | True Positive Rate |
| SOF | Start of Frame |
| RTR | Remote Transmission Request |
| IDE | Identifier Extension |
| CRC | Cyclical Redundancy Check |
| ACK | Acknowledgment |
| EOF | End of Frame |
| IFS | Inter Frame Space |
| DEL | Delimiter |
| ID | Identifier |
| OTA | Over the Air |
| DNN | Deep Neural Network |
| DBN | Deep Belief network |
| DoS | Denial of Service |
| ARP | Address Resolution Protocol |
| RPM | Radiation Portal Monitors |
| GAN | Generative Adversarial Network |
| DCAE | Deep Contractive Auto Encoder |
| DCNN | Deep Convolutional Neural Network |
| IoV | Internet of Vehicle |
| MMN | Minimum Maximum Normalization |
| MMD | Maximum Mean Discrepancy |
| HEX2DEC | Hexadecimal to Decimal |
| R-MPFR | Multiple Precision Floating-Point Reliable |
| ROC-AUC | Receiver Operating Characteristic-Area Under the Curve |